\documentclass[10pt,conference,a4paper]{IEEEtran}
\usepackage{times,amsmath,epsfig}
\usepackage{epstopdf}
\usepackage{algorithm}
\usepackage{algpseudocode}
\usepackage{color}
\usepackage{pstricks, pst-node, pst-plot, pst-circ}
\usepackage{moredefs}
\usepackage{psfrag}

\title{\LARGE Distributed Parallel Image Signal Extrapolation Framework using Message Passing Interface}

\author{ {J{\"u}rgen Seiler and Andr{\'e} Kaup}

\vspace{1.6mm}\\
\fontsize{10}{10}\selectfont\itshape
Multimedia Communications and Signal Processing\\
Friedrich-Alexander-Universit{\"a}t Erlangen-N{\"u}rnberg (FAU), Cauerstr. 7, 91058 Erlangen, Germany\\
\fontsize{9}{9}\selectfont\ttfamily\upshape
\,juergen.seiler@FAU.de,  
\,andre.kaup@FAU.de\\
}

\begin{document}
\maketitle


\begin{abstract}
This paper introduces a framework for distributed parallel image signal extrapolation. Since high-quality image signal processing often comes along with a high computational complexity, a parallel execution is desirable. The proposed framework allows for the application of existing image signal extrapolation algorithms without the need to modify them for a parallel processing. The unaltered application of existing algorithms is achieved by dividing input images into overlapping tiles which are distributed to compute nodes via Message Passing Interface. In order to keep the computational overhead low, a novel image tiling algorithm is proposed. Using this algorithm, a nearly optimum tiling is possible at a very small processing time. For showing the efficacy of the framework, it is used for parallelizing a high-complexity extrapolation algorithm. Simulation results show that the proposed framework has no negative impact on extrapolation quality while at the same time offering good scaling behavior on compute clusters.
\\[1\baselineskip]
\end{abstract}


\section{Introduction}
Looking back at the recent years, it can be discovered that the resolution of images and videos has steadily increased. This trend is very likely to persist for the coming years, introducing challenges to the algorithms and platforms that are utilized for processing this data. One of the problems that further arises is that with the high resolutions, a demand for a high quality processing comes along which typically further increases the complexity. This paper covers one class of image processing algorithms which is very important and at the same time computationally very demanding: image signal extrapolation. The task of image signal extrapolation is that an image which contains missing or undesired areas is reconstructed by extrapolating the available information into the unavailable regions. This extrapolation is required for very different tasks like error concealment, inpainting, defect pixel compensation, or resolution enhancement. In literature, there exist very different algorithms for solving this underdetermined problem as, e.g., patch-based inpainting \cite{Criminisi2004}, variational expectation-maximization based algorithms \cite{Li2008}, statistically driven modeling \cite{Koloda2014}, or the sparsity-based Frequency Selective Extrapolation (FSE) \cite{Seiler2010c}. Comparing existing extrapolation algorithms, it can be discovered that as soon as they are able to reach a high reconstruction quality, they only achieve this at the cost of a high computational complexity as shown for example \mbox{in \cite{Seiler2015}.}

This problem can be alleviated by exploiting the available compute power in the best possible way. As the single-thread performance of a central processing unit (CPU) could not be further increased that strongly, in the recent years, the trend came up to use multi-core designs for CPUs \cite{Blake2009} which follow a shared-memory model and therefore allow for a processing of several threads on the same data. The use of special devices like GPUs or the Intel Xeon Phi could be regarded as a further extension of this concept. However, algorithms always would have to be adapted to the multi-threaded operation. As an example, in \cite{Seiler2011c}, it was shown how FSE could be effectively adapted to this, yielding a high parallelization gain. 

However, the problem with shared-memory processing is that the number of CPU cores that is available in a computer cannot be arbitrarily enlarged and that the cost of computers which contain more than one CPU increases disproportionately. Thus, for further increasing the computing power, one has to switch to compute clusters. There, the CPUs do not have access to a shared memory and have to explicitly communicate among each other to solve the given task. One concept which allows for this is the Message Passing Interface (MPI) and especially its open source implementation OpenMPI \cite{Gabriel2004}. This technique is quite often used in High Performance Computing and large effort is spent for adapting algorithms to exploit the capabilities of the underlying cluster in the best possible way. However, this often results in extensive redesigns and the direct use of already available algorithms often is not possible.

In this paper, a framework is introduced which allows for the use of compute clusters for image signal extrapolation purposes without the need to redesign the underlying algorithm. The parallelization builds up on OpenMPI, but at the same time the signal extrapolation algorithms do not have to be adapted and arbitrary algorithms can be used without modification. This is achieved by dividing the input image into overlapping tiles that are distributed to the compute nodes. For this purpose, a novel image tiling algorithm is introduced which operates very fast but at the same time yields an efficient tiling of the image. Within this paper, FSE will be considered as an example algorithm since it allows for a high quality reconstruction while at the same time being computationally \mbox{demanding \cite{Seiler2010c}.}

\begin{figure*}
	\centering 
	\includegraphics[width=0.9\textwidth]{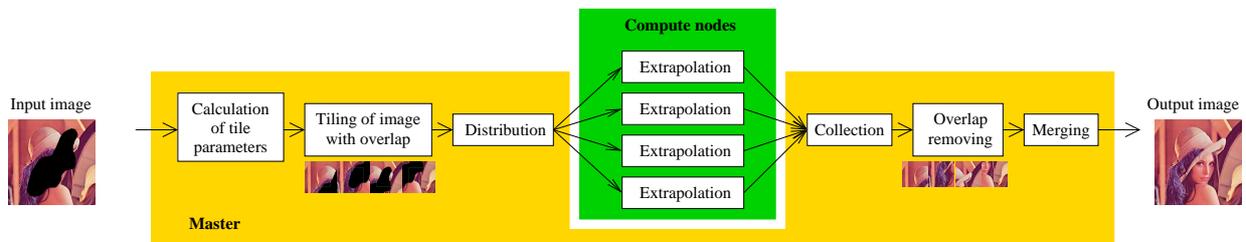}
	\caption{Extrapolation framework for parallel extrapolation of unknown areas in the input image.}
	\label{fig:extrapolation_framework}
\end{figure*}

Concepts for the parallelization of image processing tasks on clusters already exist in literature, however, they are not suited for image signal extrapolation. For example, in \cite{Squyres1995}, a small scale division is considered with focus on very simple operations on large data sets. In \cite{Afrah2008}, a middleware for vision processing is introduced at a very high abstraction level and the authors in \cite{Perrine2003} especially focus on the visualization of large amounts of data.

The paper is structured as follows. In the next section, the FSE algorithm is briefly revisited in order to provide a compact overview and to discuss the requirements of image signal extrapolation algorithms on the parallel extrapolation framework. Afterwards, the proposed distributed parallel extrapolation framework is introduced. Finally, simulation results are presented for assessing the abilities of the proposed concept, followed by the conclusion.

\section{Brief Overview of Frequency Selective Extrapolation}
The Frequency Selective Extrapolation (FSE) \cite{Seiler2010c} is a very generic signal extrapolation algorithm which also offers a high reconstruction quality. The basic idea of FSE is to divide an image into small blocks and reconstruct missing pixels within each block by generating a sparse model of the signal as weighted superposition of Fourier basis functions. To achieve the modeling, not only the known pixels of the currently considered block are used, but rather pixels from neighboring blocks as well, in order to additionally exploit this information. 

On top of this, in \cite{Seiler2011c} an algorithm is introduced for taking the shape and the distribution of the erroneous or unavailable pixels into account for deriving a useful processing order. In doing so, the algorithm is able to close large connected areas to be extrapolated from the outer margins to the center. Using this, the available information from all directions is used allowing for a stable estimation of unavailable pixels, even though the spatial distance to known pixels might be large. 

In general, FSE performs best, if the region to be extrapolated is completely surrounded by known samples. This is a property that is also true for other image signal extrapolation algorithms and all algorithms perform inferior if there is no spatial information all around the boundaries of the area to be extrapolated.

\section{OpenMPI-Parallelization Framework}
In the following, the novel parallelization framework is introduced that allows for the extrapolation of image signals on a compute cluster of independent nodes without the necessity to redesign the underlying extrapolation algorithm. The basic idea of the framework is to divide the image to be extrapolated in a set of rectangular tiles where the number of tiles is equal to the number of compute nodes. As the extrapolation quality typically is higher if the area to be extrapolated is completely enclosed by known samples, the framework uses an overlap between the tiles. In doing so, distortions by boundary effects can be alleviated and an independent processing of the tiles at the compute nodes is possible. The parallel extrapolation framework can be described as shown in Fig.\ \ref{fig:extrapolation_framework}. Starting with the input image with regions to be extrapolated, the master first calculates the parameters of the tiling. Afterwards, the image is divided into the tiles while a stripe of $d$ samples width is added to each tile to allow an overlap of the different tiles. In doing so, it is possible to consider pixels in one tile for the extrapolation in the neighboring tile, as well. In the next step, the tiles are distributed to the compute nodes via MPI. In order to avoid delays, a non-blocking communication should be used for this. All these tiles can be processed independently at the compute nodes and the unknown areas can be extrapolated there. The extrapolated tiles are collected by the master again, the overlapping regions are removed, and the tiles are merged together in order to obtain the output image.

Of course, the overlap of the tiles is necessary for having enough information available for the extrapolation, even in the case that the area to be extrapolated is located at the outer margin of the tile. However, as the overlapping regions also have to be processed, a computational overhead arises which harms the overall efficiency. Thus, it is of great importance to use an intelligent tiling in order to keep the computational overhead as small as possible. Since the number of pixels which have to be processed as overhead directly depends on the length of the boundary of the tiles, the tiling should be carried out in such a way that the tiles are preferably compact, that is to say, have a good ratio between the area and the length of the boundary. In the ideal case, this would be squares.

However, this ideal case usually cannot be assured, since the geometry of the input image and the number of tiles have to be considered. Thus, a sensible tiling algorithm is required which satisfies the following properties:
\begin{enumerate}
	\item The tiling should produce tiles of approximately equal size in order to put equal load to each compute node.
	\item The boundary of the tiles should be as small as possible in order to keep overhead caused by the overlapping regions small.
	\item The tiling algorithm itself has to exhibit a low complexity as it is running solely at the master.
\end{enumerate}
In \cite{Bezrukov1997}, an algorithm is proposed which perfectly fulfills the first two properties. This algorithms achieves an optimum tiling, however this is bought at the expense of a high computational complexity. The algorithm can be efficiently implemented recursively, but as it makes use of a brute-force search, its complexity increases exponentially with the number of tiles. Thus, the third property is not met. Alternatively, one might think of tiling an image into horizontal or vertical stripes. This could be achieved extremely efficient and a division into equally sized tiles is no problem. However, the boundary length is longer than necessary, yielding a high computational overhead.

As neither of these two algorithms can fulfill all requirements, a novel tiling algorithm is proposed in the following. This algorithm can be used within the parallel extrapolation framework presented above and provides a tiling which is nearly as good as the one presented in \cite{Bezrukov1997} while the computational complexity is of the same magnitude as for the case of a fixed vertical or horizontal division. 

The task for the proposed tiling algorithm is to divide an image of width $W$ and height $H$ into $N$ tiles. For the presentation, landscape orientation is assumed, i.e., $W>H$. If this not fulfilled, the coordinate axes have to be swapped accordingly. The first step is to calculate the aspect ratio
\begin{equation}
r = W / H
\end{equation} 
of the image. This is necessary in order to determine how many rows of tiles should be present in the image. Since the primary objective of the tiling algorithm is to obtain compact tiles that have a good ratio between area and boundary, the aspect ratio is considered for calculating the number of rows
\begin{equation}
n_R = \mathrm{max}\left(1, \mathrm{round} \left(\sqrt{N / r}\right)\right).
\end{equation} 
In this context, the rounding operation is necessary since $n_R$ obviously has to be integer.

Afterwards, the number of tiles $n_C$ in every row is determined. As $n_C$ also has to be integer, it is not possible to just divide the desired number of tiles by the number of rows, but rather they have to be distributed unequally. In order to allow for a fair distribution of tiles per row, $n_C$ is calculated as follows
\begin{equation}
n_C = \left\{ \begin{array}{ll} \lfloor N / n_R \rfloor & \mbox{for } i > N \bmod n_R \\ \lceil N / n_R \rceil &  \mbox{for } i \leq N \bmod n_R \end{array} \right.
\end{equation}
where $i$ depicts the row index. After the number of tiles in the currently considered tile row has been determined, the height $h_T$ of all tiles in this row have to be determined. Of course, this depends on the number of tiles in the row and is determined by
\begin{equation}
h_T = \lfloor H \cdot n_C / N\rfloor + \lfloor \left((H \cdot n_C) \bmod N\right) / n_R \rfloor .
\end{equation}
The height of the last row is selected so that the whole image height is covered. Next, width $w_T$ of the different tiles in the currently considered row has to be calculated. Again, it has to be considered that $w_T$ is integer. With $j$ depicting the index of the tile in the current row, 
\begin{equation}
w_T = \left\{ \begin{array}{ll} \lfloor W / n_C \rfloor & \mbox{for } j > W \bmod n_C \\ \lceil W / n_C \rceil &  \mbox{for } j \leq W \bmod n_C \end{array} \right.
\end{equation} 
follows.

\begin{figure}
\centering
\psfrag{Proposed tiling:}[l][l][1][90]{Proposed tiling}
\psfrag{Vertical tiling:}[l][l][1][90]{Vertical tiling}
\psfrag{Bezrukov tiling:}[l][l][1][90]{Bezrukov tiling}
\includegraphics[width=0.33\textwidth]{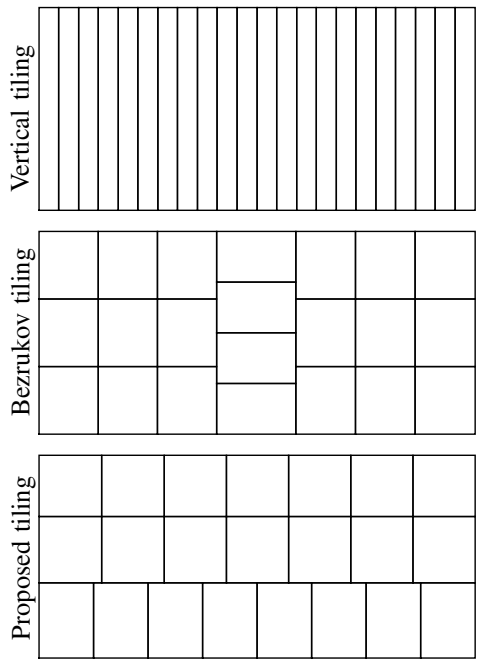}
\caption{Example for tiling an image of width $W=1024$ and height $H=480$ in $N=22$ tiles.}
\label{fig:tiling_example}
\end{figure}

By applying the above presented operations, the tiling parameters can be determined quite efficiently. In order to provide a compact overview of the proposed tiling algorithm, its pseudo code is given in Algorithm \ref{algo:tiling}. The outputs of the algorithm are on the one hand the horizontal and vertical offsets $x_T$ and $y_T$ of the individual tiles, and on the other hand the corresponding widths $w_T$ and heights $h_T$. In addition to the pseudo code, Fig.\ \ref{fig:tiling_example} shows an example for tiling an image of width $W=1024$ and height $H=480$ into $N=22$ tiles, either in vertical tiles, or using the tiling from Bezrukov \cite{Bezrukov1997}, or the proposed one. It can be discovered, that the objective of the algorithm is fulfilled and that the output tiles are compact as well as of equal size, in the same way as would be achieved by the elaborate algorithm from \cite{Bezrukov1997}.

\floatname{algorithm}{Algorithm}
\begin{algorithm}[tp]
\caption{Proposed algorithm for dividing an image into compact tiles. $x_T$ and $y_T$ depict offset of the tiles, $w_T$ and $h_T$ their width and height. The input image is assumed to be in landscape format.}
\label{algo:tiling}
\begin{algorithmic}[5]
\renewcommand{\algorithmicrequire}{\textbf{Input:}}
\renewcommand{\algorithmicensure}{\textbf{Output:}}
\Require Image width $W$, image height $H$, number of tiles $N$
\Ensure $N$ compact tiles of approximately equal area
\State Aspect ratio $r = W / H$
\State Number of rows $n_R = \mathrm{max}\left(1, \mathrm{round} \left(\sqrt{N / r}\right)\right)$
\State $x_T=0,y_T=0$
\ForAll{$i = 1, ... , n_R$}
	\State Number of tiles in current row 	\\ $\hspace{1.5cm} n_C = \left\{ \begin{array}{ll} \lfloor N / n_R \rfloor & \mbox{for } i > N \bmod n_R \\ \lceil N / n_R \rceil &  \mbox{for } i \leq N \bmod n_R \end{array} \right.$
	\State $h_T = \lfloor H \cdot n_C / N\rfloor + \lfloor \left((H \cdot n_C) \bmod N\right) / n_R \rfloor$
	\If{$i = n_R$}
		$h_T = H - y_T$
	\EndIf
	\ForAll{$j = 1, ... , n_C$}
		\State $w_T = \left\{ \begin{array}{ll} \lfloor W / n_C \rfloor & \mbox{for } j > W \bmod n_C \\ \lceil W / n_C \rceil &  \mbox{for } j \leq W \bmod n_C \end{array} \right.$
		\State Store $x_T,\ y_T,\ w_T,\ h_T$ for current tile
		\State $x_T = x_T + w_T$
	\EndFor
	\State $y_T = y_T + h_T,\ x_T=0$
\EndFor
\end{algorithmic}
\end{algorithm}

\section{Simulations and Results}
\begin{figure}
	\centering
	\includegraphics[width=0.4\textwidth]{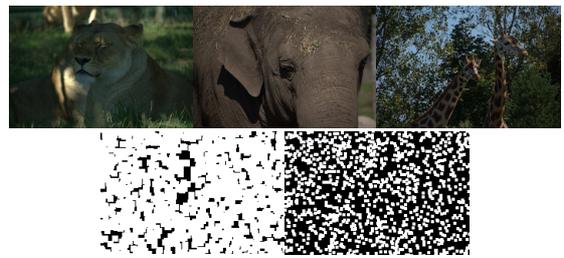}
	\caption{Test images Lion, Elephant, Giraffe and details of size $240\times 160$ pixels of masks Mask1 and Mask2 used for evaluation.}
	\label{fig:testimages}
\end{figure}

For evaluating the performance of the proposed parallel image signal extrapolation framework, tests have been carried out on a small test cluster. The master of the cluster is an Intel Core2 Quad Q9550 and for the computation four AMD Opteron 6274 are available. Each of these CPUs has $8$ independent cores. By using every core as independent node, the cluster is able to provide up to $32$ compute nodes of equal speed and a tiling of up to the same number can be tested. The cluster is running the Rocks Cluster Distribution system \cite{Rocks2015} with OpenMPI version 1.6.2. The FSE used for the extrapolation is written in C and is running single-threaded on each node. The parameters for FSE have been selected according to \cite{Seiler2011c} and a block size of $4$ is used. The used test images are shown in Fig.\ \ref{fig:testimages} and are called Lion, Elephant and Giraffe and are of size of $6024\times 4024$ pixels. Furthermore, two masks that define the areas to be extrapolated are considered. Using Mask1, $14\%$ of the pixels have to be extrapolated, for Mask2 $62\%$ of the pixels have to be extrapolated.

\begin{figure}
	\centering
	\input{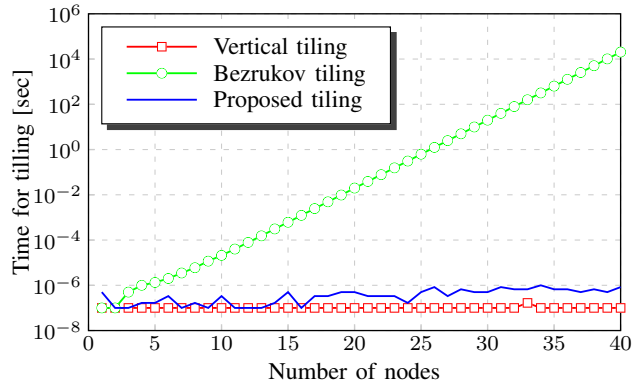}
	\caption{Processing time for calculation of tile parameters. Measurement accuracy is of the magnitude of $10^{-7}$ seconds.} 
	\label{fig:tiling_time}
\end{figure}

Since the extrapolation algorithm is left unchanged, there are two factors in the proposed parallel extrapolation framework that have an impact on the performance. This is the algorithm used for the tiling on the one hand and the size $d$ of the overlap between the tiles on the other hand. In order to analyze the first one, Fig.\ \ref{fig:tiling_time} shows the time that is necessary just for calculating the tiling parameters at the master. For this, the cases are considered that a simple vertical tiling, the algorithm from Bezrukov \cite{Bezrukov1997}, or the proposed algorithm is used. Thereby, it has to be noted that the measurement accuracy of the time is of the magnitude of $10^{-7}$ seconds. Examining the plot, it can be seen that the time necessary for the vertical and the proposed tiling is extremely small and in the order of the measurement accuracy while the calculation time of the Bezrukov tiling exponentially increases with the number of tiles to be determined. For $32$ tiles, this already ends up in $79$ seconds just for calculating the tiling parameters.

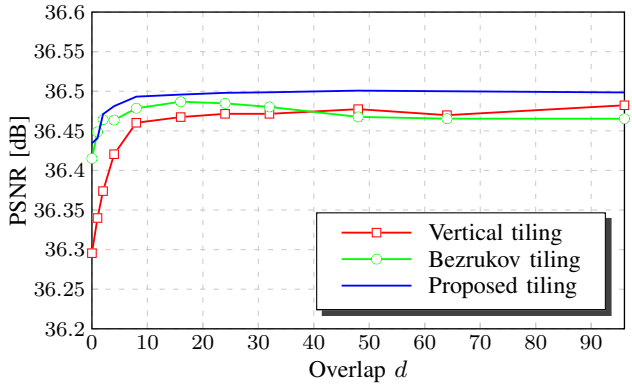
\begin{figure}
\centering



\providelength{\AxesLineWidth}       \setlength{\AxesLineWidth}{0.5pt}
\providelength{\GridLineWidth}       \setlength{\GridLineWidth}{0.4pt}
\providelength{\GridLineDotSep}      \setlength{\GridLineDotSep}{0.4pt}
\providelength{\MinorGridLineWidth}  \setlength{\MinorGridLineWidth}{0.4pt}
\providelength{\MinorGridLineDotSep} \setlength{\MinorGridLineDotSep}{0.8pt}
\providelength{\plotwidth}           \setlength{\plotwidth}{7cm} 
\providelength{\LineWidth}           \setlength{\LineWidth}{0.7pt}
\providelength{\MarkerSize}          \setlength{\MarkerSize}{4pt}
\newrgbcolor{GridColor}{0.8 0.8 0.8}

\psset{xunit=0.010417\plotwidth,yunit=1.500000\plotwidth}
\begin{pspicture}(-10.285714,36.128571)(96.000000,36.600000)


\psline[linestyle=dashed,dash=2pt 3pt,dotsep=\GridLineDotSep,linewidth=\GridLineWidth,linecolor=GridColor](0.000000,36.200000)(0.000000,36.600000)
\psline[linestyle=dashed,dash=2pt 3pt,dotsep=\GridLineDotSep,linewidth=\GridLineWidth,linecolor=GridColor](10.000000,36.200000)(10.000000,36.600000)
\psline[linestyle=dashed,dash=2pt 3pt,dotsep=\GridLineDotSep,linewidth=\GridLineWidth,linecolor=GridColor](20.000000,36.200000)(20.000000,36.600000)
\psline[linestyle=dashed,dash=2pt 3pt,dotsep=\GridLineDotSep,linewidth=\GridLineWidth,linecolor=GridColor](30.000000,36.200000)(30.000000,36.600000)
\psline[linestyle=dashed,dash=2pt 3pt,dotsep=\GridLineDotSep,linewidth=\GridLineWidth,linecolor=GridColor](40.000000,36.200000)(40.000000,36.600000)
\psline[linestyle=dashed,dash=2pt 3pt,dotsep=\GridLineDotSep,linewidth=\GridLineWidth,linecolor=GridColor](50.000000,36.200000)(50.000000,36.600000)
\psline[linestyle=dashed,dash=2pt 3pt,dotsep=\GridLineDotSep,linewidth=\GridLineWidth,linecolor=GridColor](60.000000,36.200000)(60.000000,36.600000)
\psline[linestyle=dashed,dash=2pt 3pt,dotsep=\GridLineDotSep,linewidth=\GridLineWidth,linecolor=GridColor](70.000000,36.200000)(70.000000,36.600000)
\psline[linestyle=dashed,dash=2pt 3pt,dotsep=\GridLineDotSep,linewidth=\GridLineWidth,linecolor=GridColor](80.000000,36.200000)(80.000000,36.600000)
\psline[linestyle=dashed,dash=2pt 3pt,dotsep=\GridLineDotSep,linewidth=\GridLineWidth,linecolor=GridColor](90.000000,36.200000)(90.000000,36.600000)
\psline[linestyle=dashed,dash=2pt 3pt,dotsep=\GridLineDotSep,linewidth=\GridLineWidth,linecolor=GridColor](0.000000,36.200000)(96.000000,36.200000)
\psline[linestyle=dashed,dash=2pt 3pt,dotsep=\GridLineDotSep,linewidth=\GridLineWidth,linecolor=GridColor](0.000000,36.250000)(96.000000,36.250000)
\psline[linestyle=dashed,dash=2pt 3pt,dotsep=\GridLineDotSep,linewidth=\GridLineWidth,linecolor=GridColor](0.000000,36.300000)(96.000000,36.300000)
\psline[linestyle=dashed,dash=2pt 3pt,dotsep=\GridLineDotSep,linewidth=\GridLineWidth,linecolor=GridColor](0.000000,36.350000)(96.000000,36.350000)
\psline[linestyle=dashed,dash=2pt 3pt,dotsep=\GridLineDotSep,linewidth=\GridLineWidth,linecolor=GridColor](0.000000,36.400000)(96.000000,36.400000)
\psline[linestyle=dashed,dash=2pt 3pt,dotsep=\GridLineDotSep,linewidth=\GridLineWidth,linecolor=GridColor](0.000000,36.450000)(96.000000,36.450000)
\psline[linestyle=dashed,dash=2pt 3pt,dotsep=\GridLineDotSep,linewidth=\GridLineWidth,linecolor=GridColor](0.000000,36.500000)(96.000000,36.500000)
\psline[linestyle=dashed,dash=2pt 3pt,dotsep=\GridLineDotSep,linewidth=\GridLineWidth,linecolor=GridColor](0.000000,36.550000)(96.000000,36.550000)
\psline[linestyle=dashed,dash=2pt 3pt,dotsep=\GridLineDotSep,linewidth=\GridLineWidth,linecolor=GridColor](0.000000,36.600000)(96.000000,36.600000)

\psline[linewidth=\AxesLineWidth,linecolor=GridColor](0.000000,36.200000)(0.000000,36.208000)
\psline[linewidth=\AxesLineWidth,linecolor=GridColor](10.000000,36.200000)(10.000000,36.208000)
\psline[linewidth=\AxesLineWidth,linecolor=GridColor](20.000000,36.200000)(20.000000,36.208000)
\psline[linewidth=\AxesLineWidth,linecolor=GridColor](30.000000,36.200000)(30.000000,36.208000)
\psline[linewidth=\AxesLineWidth,linecolor=GridColor](40.000000,36.200000)(40.000000,36.208000)
\psline[linewidth=\AxesLineWidth,linecolor=GridColor](50.000000,36.200000)(50.000000,36.208000)
\psline[linewidth=\AxesLineWidth,linecolor=GridColor](60.000000,36.200000)(60.000000,36.208000)
\psline[linewidth=\AxesLineWidth,linecolor=GridColor](70.000000,36.200000)(70.000000,36.208000)
\psline[linewidth=\AxesLineWidth,linecolor=GridColor](80.000000,36.200000)(80.000000,36.208000)
\psline[linewidth=\AxesLineWidth,linecolor=GridColor](90.000000,36.200000)(90.000000,36.208000)
\psline[linewidth=\AxesLineWidth,linecolor=GridColor](0.000000,36.200000)(1.152000,36.200000)
\psline[linewidth=\AxesLineWidth,linecolor=GridColor](0.000000,36.250000)(1.152000,36.250000)
\psline[linewidth=\AxesLineWidth,linecolor=GridColor](0.000000,36.300000)(1.152000,36.300000)
\psline[linewidth=\AxesLineWidth,linecolor=GridColor](0.000000,36.350000)(1.152000,36.350000)
\psline[linewidth=\AxesLineWidth,linecolor=GridColor](0.000000,36.400000)(1.152000,36.400000)
\psline[linewidth=\AxesLineWidth,linecolor=GridColor](0.000000,36.450000)(1.152000,36.450000)
\psline[linewidth=\AxesLineWidth,linecolor=GridColor](0.000000,36.500000)(1.152000,36.500000)
\psline[linewidth=\AxesLineWidth,linecolor=GridColor](0.000000,36.550000)(1.152000,36.550000)
\psline[linewidth=\AxesLineWidth,linecolor=GridColor](0.000000,36.600000)(1.152000,36.600000)

{ \footnotesize 
\rput[t](0.000000,36.192000){$0$}
\rput[t](10.000000,36.192000){$10$}
\rput[t](20.000000,36.192000){$20$}
\rput[t](30.000000,36.192000){$30$}
\rput[t](40.000000,36.192000){$40$}
\rput[t](50.000000,36.192000){$50$}
\rput[t](60.000000,36.192000){$60$}
\rput[t](70.000000,36.192000){$70$}
\rput[t](80.000000,36.192000){$80$}
\rput[t](90.000000,36.192000){$90$}
\rput[r](-1.152000,36.200000){$36.2$}
\rput[r](-1.152000,36.250000){$36.25$}
\rput[r](-1.152000,36.300000){$36.3$}
\rput[r](-1.152000,36.350000){$36.35$}
\rput[r](-1.152000,36.400000){$36.4$}
\rput[r](-1.152000,36.450000){$36.45$}
\rput[r](-1.152000,36.500000){$36.5$}
\rput[r](-1.152000,36.550000){$36.55$}
\rput[r](-1.152000,36.600000){$36.6$}
} 

\pspolygon[linewidth=\AxesLineWidth](0.000000,36.200000)(96.000000,36.200000)(96.000000,36.600000)(0.000000,36.600000)(0.000000,36.200000)

{ \small 
\rput[b](48.000000,36.128571){
\begin{tabular}{c}
Overlap $d$\\
\end{tabular}
}

\rput[t]{90}(-15.285714,36.400000){
\begin{tabular}{c}
PSNR [dB]\\
\end{tabular}
}
} 

\newrgbcolor{color233.0023}{1  0  0}
\savedata{\mydata}[{
{0.000000,36.295500},{1.000000,36.339900},{2.000000,36.373900},{4.000000,36.420500},{8.000000,36.460200},
{16.000000,36.467400},{24.000000,36.471500},{32.000000,36.471500},{48.000000,36.477400},{64.000000,36.469800},
{96.000000,36.482300}
}]
\dataplot[plotstyle=line,showpoints=true,dotstyle=Bsquare,dotsize=\MarkerSize,linestyle=solid,linewidth=\LineWidth,linecolor=color233.0023]{\mydata}

\newrgbcolor{color234.0018}{0  1  0}
\savedata{\mydata}[{
{0.000000,36.415100},{1.000000,36.448700},{2.000000,36.464200},{4.000000,36.463600},{8.000000,36.478500},
{16.000000,36.486800},{24.000000,36.484700},{32.000000,36.480200},{48.000000,36.467600},{64.000000,36.465400},
{96.000000,36.465300}
}]
\dataplot[plotstyle=line,showpoints=true,dotstyle=o,dotsize=\MarkerSize,linestyle=solid,linewidth=\LineWidth,linecolor=color234.0018]{\mydata}

\newrgbcolor{color235.0018}{0  0  1}
\savedata{\mydata}[{
{0.000000,36.434400},{1.000000,36.440600},{2.000000,36.471100},{4.000000,36.481200},{8.000000,36.493200},
{16.000000,36.495800},{24.000000,36.498000},{32.000000,36.498700},{48.000000,36.500800},{64.000000,36.500000},
{96.000000,36.498500}
}]
\dataplot[plotstyle=line,showpoints=true,dotstyle=asterisk,dotsize=\MarkerSize,linestyle=solid,linewidth=\LineWidth,linecolor=color235.0018]{\mydata}

{ \small 
\rput[br](93.696000,36.216000){%
\psshadowbox[framesep=0pt,linewidth=\AxesLineWidth]{\psframebox*{\begin{tabular}{l}
\Rnode{a1}{\hspace*{0.0ex}} \hspace*{0.7cm} \Rnode{a2}{~~Vertical tiling} \\
\Rnode{a3}{\hspace*{0.0ex}} \hspace*{0.7cm} \Rnode{a4}{~~Bezrukov tiling} \\
\Rnode{a5}{\hspace*{0.0ex}} \hspace*{0.7cm} \Rnode{a6}{~~Proposed tiling} \\
\end{tabular}}
\ncline[linestyle=solid,linewidth=\LineWidth,linecolor=color233.0023]{a1}{a2} \ncput{\psdot[dotstyle=Bsquare,dotsize=\MarkerSize,linecolor=color233.0023]}
\ncline[linestyle=solid,linewidth=\LineWidth,linecolor=color234.0018]{a3}{a4} \ncput{\psdot[dotstyle=o,dotsize=\MarkerSize,linecolor=color234.0018]}
\ncline[linestyle=solid,linewidth=\LineWidth,linecolor=color235.0018]{a5}{a6} \ncput{\psdot[dotstyle=asterisk,dotsize=\MarkerSize,linecolor=color235.0018]}
}%
}%
} 

\end{pspicture}%
\caption{Influence of overlap size $d$ on reconstruction quality for test image Elephant and Mask2.} 
\label{fig:tile_border}
\end{figure}

The second important parameter is the width $d$ of the overlapping areas. As discussed above, a proper overlap is required as otherwise boundary effects may occur which harm the overall extrapolation quality. However, the overlapping areas increase the absolute area to be processed, having an impact on the overhead for the parallelization. In order to assess the impact on the extrapolation quality, Fig. \ref{fig:tile_border} shows the reconstruction quality in terms of PSNR with respect to the width $d$ of the overlapping area for the test image Elephant and Mask2. It can be seen quite well that small values of $d$ yield a reduced extrapolation quality. However, it can also be seen that an overlap of $d=32$ pixels is sufficient and larger overlapping areas do not yield any improvement. A similar behavior can be observed for the other test images and mask, as well.

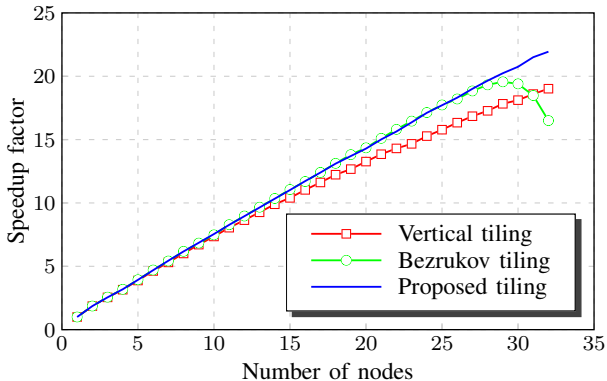
\begin{figure}
\centering



\providelength{\AxesLineWidth}       \setlength{\AxesLineWidth}{0.5pt}
\providelength{\GridLineWidth}       \setlength{\GridLineWidth}{0.4pt}
\providelength{\GridLineDotSep}      \setlength{\GridLineDotSep}{0.4pt}
\providelength{\MinorGridLineWidth}  \setlength{\MinorGridLineWidth}{0.4pt}
\providelength{\MinorGridLineDotSep} \setlength{\MinorGridLineDotSep}{0.8pt}
\providelength{\plotwidth}           \setlength{\plotwidth}{7cm} 
\providelength{\LineWidth}           \setlength{\LineWidth}{0.7pt}
\providelength{\MarkerSize}          \setlength{\MarkerSize}{4pt}
\newrgbcolor{GridColor}{0.8 0.8 0.8}

\psset{xunit=0.028571\plotwidth,yunit=0.024000\plotwidth}
\begin{pspicture}(-3.750000,-4.464286)(35.000000,25.000000)


\psline[linestyle=dashed,dash=2pt 3pt,dotsep=\GridLineDotSep,linewidth=\GridLineWidth,linecolor=GridColor](0.000000,0.000000)(0.000000,25.000000)
\psline[linestyle=dashed,dash=2pt 3pt,dotsep=\GridLineDotSep,linewidth=\GridLineWidth,linecolor=GridColor](5.000000,0.000000)(5.000000,25.000000)
\psline[linestyle=dashed,dash=2pt 3pt,dotsep=\GridLineDotSep,linewidth=\GridLineWidth,linecolor=GridColor](10.000000,0.000000)(10.000000,25.000000)
\psline[linestyle=dashed,dash=2pt 3pt,dotsep=\GridLineDotSep,linewidth=\GridLineWidth,linecolor=GridColor](15.000000,0.000000)(15.000000,25.000000)
\psline[linestyle=dashed,dash=2pt 3pt,dotsep=\GridLineDotSep,linewidth=\GridLineWidth,linecolor=GridColor](20.000000,0.000000)(20.000000,25.000000)
\psline[linestyle=dashed,dash=2pt 3pt,dotsep=\GridLineDotSep,linewidth=\GridLineWidth,linecolor=GridColor](25.000000,0.000000)(25.000000,25.000000)
\psline[linestyle=dashed,dash=2pt 3pt,dotsep=\GridLineDotSep,linewidth=\GridLineWidth,linecolor=GridColor](30.000000,0.000000)(30.000000,25.000000)
\psline[linestyle=dashed,dash=2pt 3pt,dotsep=\GridLineDotSep,linewidth=\GridLineWidth,linecolor=GridColor](35.000000,0.000000)(35.000000,25.000000)
\psline[linestyle=dashed,dash=2pt 3pt,dotsep=\GridLineDotSep,linewidth=\GridLineWidth,linecolor=GridColor](0.000000,0.000000)(35.000000,0.000000)
\psline[linestyle=dashed,dash=2pt 3pt,dotsep=\GridLineDotSep,linewidth=\GridLineWidth,linecolor=GridColor](0.000000,5.000000)(35.000000,5.000000)
\psline[linestyle=dashed,dash=2pt 3pt,dotsep=\GridLineDotSep,linewidth=\GridLineWidth,linecolor=GridColor](0.000000,10.000000)(35.000000,10.000000)
\psline[linestyle=dashed,dash=2pt 3pt,dotsep=\GridLineDotSep,linewidth=\GridLineWidth,linecolor=GridColor](0.000000,15.000000)(35.000000,15.000000)
\psline[linestyle=dashed,dash=2pt 3pt,dotsep=\GridLineDotSep,linewidth=\GridLineWidth,linecolor=GridColor](0.000000,20.000000)(35.000000,20.000000)
\psline[linestyle=dashed,dash=2pt 3pt,dotsep=\GridLineDotSep,linewidth=\GridLineWidth,linecolor=GridColor](0.000000,25.000000)(35.000000,25.000000)

\psline[linewidth=\AxesLineWidth,linecolor=GridColor](0.000000,0.000000)(0.000000,0.500000)
\psline[linewidth=\AxesLineWidth,linecolor=GridColor](5.000000,0.000000)(5.000000,0.500000)
\psline[linewidth=\AxesLineWidth,linecolor=GridColor](10.000000,0.000000)(10.000000,0.500000)
\psline[linewidth=\AxesLineWidth,linecolor=GridColor](15.000000,0.000000)(15.000000,0.500000)
\psline[linewidth=\AxesLineWidth,linecolor=GridColor](20.000000,0.000000)(20.000000,0.500000)
\psline[linewidth=\AxesLineWidth,linecolor=GridColor](25.000000,0.000000)(25.000000,0.500000)
\psline[linewidth=\AxesLineWidth,linecolor=GridColor](30.000000,0.000000)(30.000000,0.500000)
\psline[linewidth=\AxesLineWidth,linecolor=GridColor](35.000000,0.000000)(35.000000,0.500000)
\psline[linewidth=\AxesLineWidth,linecolor=GridColor](0.000000,0.000000)(0.420000,0.000000)
\psline[linewidth=\AxesLineWidth,linecolor=GridColor](0.000000,5.000000)(0.420000,5.000000)
\psline[linewidth=\AxesLineWidth,linecolor=GridColor](0.000000,10.000000)(0.420000,10.000000)
\psline[linewidth=\AxesLineWidth,linecolor=GridColor](0.000000,15.000000)(0.420000,15.000000)
\psline[linewidth=\AxesLineWidth,linecolor=GridColor](0.000000,20.000000)(0.420000,20.000000)
\psline[linewidth=\AxesLineWidth,linecolor=GridColor](0.000000,25.000000)(0.420000,25.000000)

{ \footnotesize 
\rput[t](0.000000,-0.500000){$0$}
\rput[t](5.000000,-0.500000){$5$}
\rput[t](10.000000,-0.500000){$10$}
\rput[t](15.000000,-0.500000){$15$}
\rput[t](20.000000,-0.500000){$20$}
\rput[t](25.000000,-0.500000){$25$}
\rput[t](30.000000,-0.500000){$30$}
\rput[t](35.000000,-0.500000){$35$}
\rput[r](-0.420000,0.000000){$0$}
\rput[r](-0.420000,5.000000){$5$}
\rput[r](-0.420000,10.000000){$10$}
\rput[r](-0.420000,15.000000){$15$}
\rput[r](-0.420000,20.000000){$20$}
\rput[r](-0.420000,25.000000){$25$}
} 

\pspolygon[linewidth=\AxesLineWidth](0.000000,0.000000)(35.000000,0.000000)(35.000000,25.000000)(0.000000,25.000000)(0.000000,0.000000)

{ \small 
\rput[b](17.500000,-4.464286){
\begin{tabular}{c}
Number of nodes\\
\end{tabular}
}

\rput[t]{90}(-3.750000,12.500000){
\begin{tabular}{c}
Speedup factor\\
\end{tabular}
}
} 

\newrgbcolor{color233.0062}{1  0  0}
\savedata{\mydata}[{
{1.000000,1.000000},{2.000000,1.858674},{3.000000,2.547585},{4.000000,3.154313},{5.000000,3.879350},
{6.000000,4.630346},{7.000000,5.308752},{8.000000,6.012960},{9.000000,6.710984},{10.000000,7.354825},
{11.000000,8.047979},{12.000000,8.642542},{13.000000,9.262526},{14.000000,9.897287},{15.000000,10.401451},
{16.000000,11.032349},{17.000000,11.617282},{18.000000,12.229503},{19.000000,12.670371},{20.000000,13.265584},
{21.000000,13.846647},{22.000000,14.305245},{23.000000,14.665616},{24.000000,15.276913},{25.000000,15.782764},
{26.000000,16.337026},{27.000000,16.845690},{28.000000,17.272795},{29.000000,17.826969},{30.000000,18.113942},
{31.000000,18.594480},{32.000000,19.017451}
}]
\dataplot[plotstyle=line,showpoints=true,dotstyle=Bsquare,dotsize=\MarkerSize,linestyle=solid,linewidth=\LineWidth,linecolor=color233.0062]{\mydata}

\newrgbcolor{color234.0057}{0  1  0}
\savedata{\mydata}[{
{1.000000,1.000000},{2.000000,1.863774},{3.000000,2.564970},{4.000000,3.184975},{5.000000,3.942170},
{6.000000,4.706593},{7.000000,5.416706},{8.000000,6.182690},{9.000000,6.832042},{10.000000,7.484439},
{11.000000,8.286957},{12.000000,8.968632},{13.000000,9.665509},{14.000000,10.354158},{15.000000,11.060306},
{16.000000,11.691870},{17.000000,12.410321},{18.000000,13.126669},{19.000000,13.814698},{20.000000,14.350343},
{21.000000,15.098863},{22.000000,15.809042},{23.000000,16.458092},{24.000000,17.137057},{25.000000,17.735298},
{26.000000,18.207460},{27.000000,18.841048},{28.000000,19.323169},{29.000000,19.562825},{30.000000,19.400428},
{31.000000,18.479100},{32.000000,16.500513}
}]
\dataplot[plotstyle=line,showpoints=true,dotstyle=o,dotsize=\MarkerSize,linestyle=solid,linewidth=\LineWidth,linecolor=color234.0057]{\mydata}

\newrgbcolor{color235.0057}{0  0  1}
\savedata{\mydata}[{
{1.000000,1.000000},{2.000000,1.860726},{3.000000,2.550358},{4.000000,3.187925},{5.000000,3.931410},
{6.000000,4.692329},{7.000000,5.442315},{8.000000,6.173889},{9.000000,6.858054},{10.000000,7.547188},
{11.000000,8.271641},{12.000000,8.953512},{13.000000,9.642225},{14.000000,10.335886},{15.000000,11.020103},
{16.000000,11.724667},{17.000000,12.399631},{18.000000,13.102854},{19.000000,13.702091},{20.000000,14.290164},
{21.000000,15.015087},{22.000000,15.639628},{23.000000,16.362801},{24.000000,17.111503},{25.000000,17.711267},
{26.000000,18.310360},{27.000000,19.010004},{28.000000,19.651673},{29.000000,20.239193},{30.000000,20.749299},
{31.000000,21.503482},{32.000000,21.939823}
}]
\dataplot[plotstyle=line,showpoints=true,dotstyle=asterisk,dotsize=\MarkerSize,linestyle=solid,linewidth=\LineWidth,linecolor=color235.0057]{\mydata}

{ \small 
\rput[br](34.160000,1.000000){%
\psshadowbox[framesep=0pt,linewidth=\AxesLineWidth]{\psframebox*{\begin{tabular}{l}
\Rnode{a1}{\hspace*{0.0ex}} \hspace*{0.7cm} \Rnode{a2}{~~Vertical tiling} \\
\Rnode{a3}{\hspace*{0.0ex}} \hspace*{0.7cm} \Rnode{a4}{~~Bezrukov tiling} \\
\Rnode{a5}{\hspace*{0.0ex}} \hspace*{0.7cm} \Rnode{a6}{~~Proposed tiling} \\
\end{tabular}}
\ncline[linestyle=solid,linewidth=\LineWidth,linecolor=color233.0062]{a1}{a2} \ncput{\psdot[dotstyle=Bsquare,dotsize=\MarkerSize,linecolor=color233.0062]}
\ncline[linestyle=solid,linewidth=\LineWidth,linecolor=color234.0057]{a3}{a4} \ncput{\psdot[dotstyle=o,dotsize=\MarkerSize,linecolor=color234.0057]}
\ncline[linestyle=solid,linewidth=\LineWidth,linecolor=color235.0057]{a5}{a6} \ncput{\psdot[dotstyle=asterisk,dotsize=\MarkerSize,linecolor=color235.0057]}
}%
}%
} 

\end{pspicture}%
\caption{Speedup of the reconstruction process with respect to the number of nodes for test images with Mask2.} 
\label{fig:speedup}
\end{figure}

Finally, Fig.\ \ref{fig:speedup} shows the speedup compared to a single-threaded operation on one node. Here, an overlap of $d=32$ is used and again the three different tiling modes are considered. It can be seen quite well that the vertical tiling scales less well. This is due to the fact that the vertical tiling produces an unnecessary large overhead. However, it can also be discovered that the tiling of the proposed algorithm is as good as the one from the optimum algorithm \cite{Bezrukov1997} and they scale equally for small number of nodes. If the number of nodes increases, the required time for calculating the tiles becomes significant for the Bezrukov tiling, harming the overall performance. For $32$ nodes, the performance even drops below the vertical tiling. All together, the proposed tiling allows a speedup of a factor of $22$ if all nodes are used. Of course, the speedup is smaller than the number of nodes, however, this is due to different reasons. First, the actual area to be processed is larger due to the overlapping tiles. Second, additional time is required for tiling the image, distributing the tiles to the nodes and collecting and combining the processed tiles again. And finally, since the number of samples to be extrapolated may vary from tile to tile, the processing time per node varies accordingly. However, since the overall time is determined by the longest running node, this further reduces the parallelization gain.

Nevertheless, it has to be noted that the actual extrapolation algorithm can be left unchanged allowing for the use of existing algorithms without the need to adapt them to a parallel processing. For this, a speedup of a factor of $22$ still is very beneficial. With smaller overlap $d$, a higher parallelization gain would be possible, but at the cost of small quality impairments. In order to show that the proposed parallel extrapolation framework has no negative impact on the extrapolation quality, Tab.\ \ref{tab:quality} shows the difference in quality, measured in PSNR and SSIM for the case that the proposed parallel framework is used, compared to an extrapolation of the whole images without tiling. Apparently, the difference is negligible and the framework exhibits no negative impact on the extrapolation quality, instead sometimes it may even lead to small gains.

\begin{table}
\caption{Loss in PSNR / SSIM caused by usage of parallel framework with $32$ nodes compared to direct extrapolation.}
\centering
	 \setlength{\tabcolsep}{1.5mm}
	 \renewcommand{\arraystretch}{1.3}
\begin{tabular}{|l|c|c|}
\hline  & Mask1 & Mask2 \\
\hline 	Elephant  & $\hphantom{-}0.0041\ \mathrm{dB}\ /\ -7.6\cdot 10^{-7} $  & $-0.0201\ \mathrm{dB}\ /\ -3.5\cdot 10^{-5} $ \\ 
\hline 	Lion  & $-0.0009\ \mathrm{dB}\ /\ 1.5\cdot 10^{-6} $ & $\hphantom{-}0.0046\ \mathrm{dB}\ /\ -3.1\cdot 10^{-7} $ \\ 
\hline 	Giraffe  & $-0.0023\ \mathrm{dB}\ /\ 1.6\cdot 10^{-7}  $ & $-0.0090\ \mathrm{dB}\ /\ -5.4\cdot 10^{-6}  $ \\ 
\hline
\end{tabular} \vspace{-5mm}
\label{tab:quality}
\end{table}

\section{Conclusions}
In this paper, a novel framework for distributed parallel image processing with application to signal extrapolation is introduced. Using this framework, the underlying extrapolation algorithms can be left unchanged and the parallelization is achieved by dividing the image into overlapping tiles. For this purpose, a novel tiling algorithm is introduced which is able to effectively divide the image into compact tiles and therewith keeping the overhead small. Tests have shown that a speedup of a factor of $22$ is possible for the case that $32$ nodes are used. At the same time, there is no negative impact on the extrapolation quality compared to the case that the underlying algorithm was directly applied to the whole image. Further research aims at including a load balancing to the tiling algorithm for making full use of heterogeneous clusters. Additionally, the distributed parallel framework will be extended to other signal processing tasks like denoising, image rendering, or super-resolution.




\begin{thebibliography}{10}
	\providecommand{\url}[1]{#1}
	\csname url@samestyle\endcsname
	\providecommand{\newblock}{\relax}
	\providecommand{\bibinfo}[2]{#2}
	\providecommand{\BIBentrySTDinterwordspacing}{\spaceskip=0pt\relax}
	\providecommand{\BIBentryALTinterwordstretchfactor}{4}
	\providecommand{\BIBentryALTinterwordspacing}{\spaceskip=\fontdimen2\font plus
		\BIBentryALTinterwordstretchfactor\fontdimen3\font minus
		\fontdimen4\font\relax}
	\providecommand{\BIBforeignlanguage}[2]{{%
			\expandafter\ifx\csname l@#1\endcsname\relax
			\typeout{** WARNING: IEEEtran.bst: No hyphenation pattern has been}%
			\typeout{** loaded for the language `#1'. Using the pattern for}%
			\typeout{** the default language instead.}%
			\else
			\language=\csname l@#1\endcsname
			\fi
			#2}}
	\providecommand{\BIBdecl}{\relax}
	\BIBdecl
	
	\bibitem{Criminisi2004}
	A.~Criminisi, P.~P{\'e}rez, and K.~Toyama, ``Region filling and object removal
	by exemplar-based image inpainting,'' \emph{IEEE Transactions on Image
		Processing}, vol.~13, no.~9, pp. 1200--1212, Sept. 2004.
	
	\bibitem{Li2008}
	X.~Li, ``Variational bayesian image processing on stochastic factor graphs,''
	in \emph{Proceedings International Conference on Image Processing}, San
	Diego, USA, Oct. 2008, pp. 1748--1751.
	
	\bibitem{Koloda2014}
	J.~Koloda, A.~M. Peinado, and V.~S{\'a}nchez, ``Kernel-based {MMSE} multimedia
	signal reconstruction and its application to spatial error concealment,''
	\emph{IEEE Transactions on Multimedia}, vol.~16, no.~6, pp. 1729--1738, Oct.
	2014.
	
	\bibitem{Seiler2010c}
	J.~Seiler and A.~Kaup, ``Complex-valued frequency selective extrapolation for
	fast image and video signal extrapolation,'' \emph{IEEE Signal Processing
		Letters}, vol.~17, no.~11, pp. 949--952, Nov. 2010.
	
	\bibitem{Seiler2015}
	J.~Seiler, M.~Jonscher, M.~Sch{\"o}berl, and A.~Kaup, ``Resampling images to a
	regular grid from a non-regular subset of pixel positions using frequency
	selective reconstruction,'' \emph{IEEE Transactions on Image Processing},
	vol.~24, no.~11, pp. 4540--4555, Nov. 2015.
	
	\bibitem{Blake2009}
	G.~Blake, R.~G. Dreslinski, and T.~Mudge, ``A survey of multicore processors,''
	\emph{IEEE Signal Processing Magazine}, vol.~26, no.~6, pp. 26--37, Nov.
	2009.
	
	\bibitem{Seiler2011c}
	J.~Seiler and A.~Kaup, ``Optimized and parallelized processing order for
	improved frequency selective signal extrapolation,'' in \emph{Proceedings
		European Signal Processing Conference}, Barcelona, Spain, Aug. 2011, pp.
	269--273.
	
	\bibitem{Gabriel2004}
	E.~Gabriel, G.~E. Fagg, G.~Bosilca, T.~Angskun, J.~J. Dongarra, J.~M. Squyres,
	V.~Sahay, P.~Kambadur, B.~Barrett, A.~Lumsdaine, R.~H. Castain, D.~J. Daniel,
	R.~L. Graham, and T.~S. Woodall, ``Open {MPI}: Goals, concept, and design of
	a next generation mpi implementation,'' in \emph{Proceedings 11th European
		PVM/MPI Users’ Group Meeting}, Budapest, Hungary, Sept. 2004, pp. 97--104.
	
	\bibitem{Squyres1995}
	J.~M. Squyres, A.~Lumsdaine, and R.~L. Stevenson, ``A cluster-based parallel
	image processing toolkit,'' in \emph{Proceedings IS\&T Conference on Image
		and Video Processing}, San Jose, USA, Feb. 1995.
	
	\bibitem{Afrah2008}
	A.~Afrah, G.~Miller, D.~Parks, M.~Finke, and S.~Fels, ``Hive: A distributed
	system for vision processing,'' in \emph{Proceedings ACM/IEEE International
		Conference on Distributed Smart Cameras}, Stanford, USA, Sept. 2008, pp.
	1--9.
	
	\bibitem{Perrine2003}
	K.~A. Perrine and D.~R. Jones, ``Interactive imaging science on parallel
	computers: getting immediate results,'' in \emph{Proceedings International
		Parallel and Distributed Processing Symposium}, Nice, France, April 2003.
	
	\bibitem{Bezrukov1997}
	S.~Bezrukov and B.~Rovan, ``On partitioning grids into equal parts,''
	\emph{Computers and Artificial Intelligence}, vol.~16, pp. 153--165, 1997.
	
	\bibitem{Rocks2015}
	\BIBentryALTinterwordspacing
	[Online]. Available: \url{http://www.rocksclusters.org/}
	\BIBentrySTDinterwordspacing
	
\end{thebibliography}
\end{document}